\documentclass[12pt,a4paper,english,superscriptaddress,aps,nofootinbib]{revtex4}
\usepackage[utf8]{inputenc}
\usepackage[T1]{fontenc}
\usepackage{amsmath,amssymb,graphicx}
\makeatletter
\usepackage{babel}
\usepackage[active]{srcltx}
\usepackage{graphicx,color}
\usepackage{changebar}
\usepackage{hyperref}
\usepackage[T1]{fontenc}
\usepackage{esint}
\usepackage{multirow}
\usepackage{dsfont}
\usepackage{ae}
\usepackage{amsmath}
\usepackage{braket}
\usepackage{mathtools}
\usepackage{slashed}
\usepackage{empheq}
\usepackage{multirow}
\usepackage{graphicx}
\usepackage{amsfonts}
\usepackage{amsmath}
\newcommand{\be}{\begin{equation}}
	\newcommand{\ee}{\end{equation}}
\newcommand{\bea}{\begin{eqnarray}}
	\newcommand{\eea}{\end{eqnarray}}

\begin{document}
	\title{Differential configurational entropy for multi-field of the $\phi^6$ theory}
	
\author{F. C. E. Lima}
\email{E-mail: cleiton.estevao@fisica.ufc.br}
\affiliation{Universidade Federal do Cear\'{a} (UFC), Departamento do F\'{i}sica - Campus do Pici, Fortaleza, CE, C. P. 6030, 60455-760, Brazil.}
 
\author{C. A. S. Almeida}
\email{E-mail: carlos@fisica.ufc.br}
\affiliation{Universidade Federal do Cear\'{a} (UFC), Departamento do F\'{i}sica - Campus do Pici, Fortaleza, CE, C. P. 6030, 60455-760, Brazil.}

\begin{abstract}
The topological structures of a $\phi^6$ theory with multi-field are studied. The $\phi^6$ theory is interesting because it is a theory that allows the shrinkage of topological structures generating double-kink or even multi-kink configurations. In this work, we consider and study the solutions of a two real scalar fields model. To reach our purpose, we investigate the BPS properties of the fields using the approach proposed by Bogomol'nyi-Prasad-Sommerfield. Using the BPS energy density, the differential configurational entropy (DCE) of the BPS structures is studied. The result of the DCE indicates the most likely field configuration of one of the topological sectors of the model.
\end{abstract}

\maketitle

\thispagestyle{empty}

\section{Introduction}

Since Finkelstein's seminal paper on kinks \cite{Finkelstein}, several works have arisen in the literature discussing these configurations \cite{G1,G2,B1,B2,A1,A2,LGA,LDA}. The applications of kink structures appear in various physical scenarios such as condensed matter theories \cite{Jubert,Vanhaverbeke,Uchida}, high energies \cite{LPA,Matsunami,Dutta}, and nonlinear theories \cite{Tin}. Recently, attractive studies on kinks in the gravitational background have aroused the interest of some researchers \cite{Zhong0}. Indeed, these studies can help to explain questions of quantum gravity \cite{Henneaux,Alwis}, gravitational collapse \cite{Vaz,Vaz1}, and black hole evaporation \cite{Callan,Bilal}.

It is incontestable that spontaneous symmetry breaking is ubiquitous in physics \cite{FCECAS}. In studies of topological structures, spontaneous symmetry breaking is essential since it relates to the phase transitions and the generation of localized structures \cite{LA1,LA2}. These topological structures are finite energy configurations and can generally appear in one, two, and three spatial dimensions. Therefore, the kinks are the topological structures that emerge in 2D theories.

The fact is that solitons in classical and quantum field theories play an important role due to their applications \cite{Manton}. In the 2D model, the kink is a solitary wave configuration \cite{Rajaraman}. This configuration interpolates between different minima of the theory. The interaction between structures such as kink-kink or kink-antikink is a research branch with increasing interest. That interest occurs due to their applications, the study of kink-antikink creation from particles (or radiation) \cite{Tomasz,Sourish} and multi-kink collisions \cite{Marjaneh,Marjaneh1} are some examples. 

Currently, the $\phi^4$ theory is the most used model in the literature, see e. g., Refs. \cite{Yan,Evslin}. In fact, this is because the $\phi^4$ model forms the phenomenological basis of phase transitions in the Ginzburg-Landau theory \cite{Saxena}. This model is interesting because it describes a symmetrical double potential well. On the other hand, higher-order theories like the $\phi^6$ theory become more interesting because they allow the model to admit multiple phase transitions. We found in the literature some studies about the $\phi^6$ model. For example, the $\phi^6$ theory with $O(N)$ symmetry, was used recently to study different phase transitions \cite{Herzog}. The $\phi^6$ models also were studied in the context of supersymmetry \cite{Kwon} and in the study of scattering of structures \cite{azadeh}.

Configurational Entropy (CE) is an important tool that helps understand the topological field configurations. CE is an extension of Shannon's theory \cite{Gleiser0,Gleiser,Shannon}. However, CE and its variants (differential configuration entropy and differential configuration complexity) are applied to continuous systems and provide information about the stability of localized structures \cite{LA1}. In other words, while Shannon's theory studies the hidden information in a random process \cite{Shannon}, CE is interpreted as a theoretical measure of the stability of the localized structures \cite{Gleiser0,Gleiser,Gleiser2,Gleiser3,Gleiser4}. The fact is that configurational entropy is applied to study several systems. For example, Gleiser et al. used CE to study Q-balls solutions \cite{Gleiser2,Gleiser4} and non-equilibrium dynamics of spontaneous symmetry breaking \cite{Gleiser4}. Meantime, Rocha et al. used CE to study black holes \cite{NR} and holographic models \cite{R}.

In this letter, we seek to understand how topological structures change in a model with two real scalar fields. In principle, in a model governed by a single scalar field with the interaction of the type $\phi^6$, one can have multiple domain walls so that the structure can take the form of a multi-kink. If the model has two scalar fields with an interaction type $\phi^6$, is it possible to have the emergence of multi-kink? That is the question that we seek to discuss throughout this work.

We organize this letter as follows: Initially, we investigate the field configurations using the BPS formalism for an arbitrary potential. Then, a particular model of the $\phi^6$ theory is exposed, and the analytical solutions of the scalar fields are displayed. Posteriorly, is investigated the field configuration more likely using the DCE concept. We finish our discussion by announcing the findings.

\section{The model}

In this paper, we propose to study the kink-like topological structures in a generalized theory in which we have two generalized scalar fields. This model has already been used by Bazeia et al. \cite{Bazeia1} to study the existence of geometrically contracted structures in the $\phi^4$ theory. In this work, let us use a similar action, although with a $\phi^6$ interaction. Therefore, allow us to assume a flat two-dimensional spacetime. In this way, the action is
\begin{align}\label{eq1}
    S=\int\, d^2x\, \bigg[\frac{1}{2}f(\chi)\partial_\mu\phi\partial^\mu\phi+\frac{1}{2}\partial_\mu\chi\partial^\mu\chi-V(\phi,\chi)\bigg].
\end{align}
The function $f(\chi)$ is the generalization function and is positively defined. In this work, we will assume the metric signature as $g_{\mu\nu}=$diag$(+,-)$. In natural units, there is still that $\hbar=c=1$. Here, it is important to highlight that the potential $V(\phi,\chi)$ couples the scalar fields at the energy saturation limit. Consequently, the interaction theory will be related to the $f(\chi)$ generalization function.

The equations of motion arise from varying the action concerning the fields $\phi$ and $\chi$. Therefore, the equation of motion of the $\phi$ field configuration is
\begin{align}
    \frac{d}{dx}\bigg(f(\chi)\frac{d\phi}{dx}\bigg)-V_\phi(\phi,\chi)=0,
\end{align}
and, the variation concerning the field $\chi$ gives us the equation of motion, namely,
\begin{align}
    \frac{d^2\chi}{dx^2}-\frac{1}{2}f_\chi\bigg(\frac{d\phi}{dx}\bigg)^2-V_\chi(\phi,\chi)=0,
\end{align}
where $V_\phi=\partial V/\partial\phi$, $V_\chi=\partial V/\partial\chi$, and $f_\chi=df/d\chi$.

To obtain kink-like topological structures, the topological conditions of the scalar fields are
\begin{align}
    \phi(x\to\pm\infty)=\nu_\pm \hspace{0.25cm} \text{and} \hspace{0.25cm} \chi(x\to\pm\infty)=u_{\pm}.
\end{align}
Here, $\nu_\pm$ and $u_\pm$ correspond to the theory's vacuum expected value (VEV).

Let us implement the Bogomol'nyi approach. For this, it is necessary to construct the energy-momentum tensor. Varying the action concerning the metric, we obtain that the energy-momentum tensor is
\begin{align}\label{Tmunu}
    T_{\mu\nu}=f(\chi)\partial_\mu\phi\partial_\nu\phi+\partial_{\mu}\chi\partial_\nu\chi-g_{\mu\nu}\mathcal{L},
\end{align}
where $\mathcal{L}$ is the Lagrangian density associated with the action (\ref{eq1}). In other words, the Lagrangian density is the integrand of the expression (\ref{eq1}).

The energy density of the fields is the $T_{00}$ component of the energy-momentum tensor. Thereby allows us to write the energy density of the structures as
\begin{align}
    \mathcal{E}=T_{00}=\frac{1}{2}f(\chi)\bigg(\frac{d\phi}{dx}\bigg)^2+\frac{1}{2}\bigg(\frac{d\chi}{dx}\bigg)^2+V(\phi,\chi).
\end{align}

Let us, at this point, turn our attention to the study of self-dual structures. For this, we will use the Bogomol'nyi-Prasad-Sommerfield (BPS) approach to restructure the energy density and find the self-dual equations of the model. Now, the energy density can be rewritten as
\begin{align}
    \mathcal{E}=&\frac{f(\chi)}{2}\bigg[\bigg(\frac{d\phi}{dx}\bigg)\mp\frac{\mathcal{W}_\phi}{f(\chi)}\bigg]^2+\frac{1}{2}\bigg[\frac{d\chi}{dx}\mp\mathcal{W}_\chi\bigg]^2+V(\phi,\chi)-\frac{\mathcal{W}_{\phi}^{2}}{2f(\chi)}-\frac{W_{\chi}^{2}}{2}\pm\frac{d\mathcal{W}}{dx}.
\end{align}
To inspect the BPS structures, we rewrite the potential in terms of an arbitrary function, i. e., the $\mathcal{W}$ function. In  Refs. \cite{LA,Vachaspati}, to study topological structures with BPS properties an auxiliary function $\mathcal{W}$ is considered. In this context, the helper function $\mathcal{W}$ is called the superpotential function \cite{Vachaspati}. In general, the implementation of this auxiliary function allows writing the potential and energy in terms of the superpotential. This approach using an auxiliary function is useful in studying the first-order formalism of topological structures in various scenarios \cite{MLSA,Zhong}.

Perceive that the energy density is limited. So, at the energy saturation limit, we obtain the BPS equations, namely,
\begin{align}
    \label{BPS1}
    \frac{d\phi}{dx}=\pm\frac{\mathcal{W}_\phi}{f(\chi)},
\end{align}
and
\begin{align}
    \label{BPS2}
    \frac{d\chi}{dx}=\pm\mathcal{W}_\chi.
\end{align}

At the energy saturation boundary, for the system to have BPS property, it is convenient to choose the model potential as
\begin{align}
    \label{potential}
    V(\phi,\chi)=\frac{\mathcal{W}_{\phi}^{2}}{2f(\chi)}+\frac{\mathcal{W}_{\chi}^{2}}{2}.
\end{align}
For this condition, the energy density of the fields is the BPS energy density, i. e.,
\begin{align}
    \mathcal{E}=\mathcal{E}_{BPS}=\pm\frac{d\mathcal{W}}{dx},
\end{align}
so that the energy of the fields at the BPS boundary is
\begin{align}
    E_{BPS}=&\int\, dx\, \mathcal{E}_{BPS}=\pm\int_{-\infty}^{\infty}\, \frac{d\mathcal{W}}{dx}\,dx=\pm[\mathcal{W}(v_{+},u_{+})-\mathcal{W}(v_{-},u_{-})],
\end{align}
i. e., 
\begin{align}\label{energy1}
    E_{BPS}=\pm[\mathcal{W}(v_{+},u_{+})-\mathcal{W}(v_{-},u_{-})].&
\end{align}

\subsection{The $\phi^6$ theory}

Before assuming a form for the auxiliary function $\mathcal{W}(\phi,\chi)$, it is important to mention that the case where first-order equations are simplest is when $\mathcal{W}_{ \phi\chi}=\mathcal{W}_{\chi\phi}=0$. In this case, an interesting particularly choice is
 \begin{align}
     \mathcal{W}(\phi,\chi)=\mathcal{W}_1(\phi)+\mathcal{W}_2(\chi).
 \end{align}

To work in a $\phi^6$ theory, let us suppose that the superpotential has the following profile,
\begin{align}\label{choose}
    \mathcal{W}(\phi,\chi)=\frac{\phi^2}{2}\bigg(1-\frac{\phi^2}{2}\bigg)+\frac{\chi^2}{2}\bigg(1-\frac{\chi^2}{2}\bigg).
\end{align}
The auxiliary function (\ref{choose}) profile is interesting because it reproduces a symmetric $\phi^6$ theory. Indeed, that superpotential is an extension of the model discussed in reference \cite{Bazeia1}. In Ref. \cite{Bazeia1}, the authors choose the superpotential for reproducing the $\phi^4$ theory. The choice (\ref{choose}) leads us to an extension of the $\phi^4$ theory, i. e., we will have a model with three distinct vacuum values. Furthermore, for this extension (\ref{choose}), it is common to expect the appearance of geometrically contracted structures, such as the class of multi-kink solutions. Motivated by this hypothesis, we assume the superpotential (\ref{choose}) and write the interaction as
\begin{align}\label{potential1}
    V(\phi,\chi)=\frac{\phi^2(1-\phi^2)^2}{2f(\chi)}+\frac{\chi^2(1-\chi^2)^2}{2}.
\end{align}
Although, in this paper, we have assumed the superpotential to produce a specific profile of the interaction. In some models, it is usual to apply the correspondences with the supersymmetric theories to study the BPS property. For these cases,  the supersymmetry (SUSY) can adjust the action to satisfy the BPS approach. In this way, SUSY can propose a profile for the auxiliary function, see Ref. \cite{CCA}. However, in our model, the study of the SUSY correspondence for the BPS approach is not applied. That's because, in principle, a polynomial function that spontaneously breaks symmetry will preserve the BPS property in our model.

Furthermore, note that the potential described above is equivalent to a theory like $\phi^6$. The $\phi^6$-theories are discussed extensively in Refs. \cite{Saxena,Herzog,azadeh}. For example, Saxena {\it et al.} \cite{Saxena} establishes the theory $\phi^6$ in a Ginzburg-Landau theory of successive phase transitions. In that work, the authors use the $\phi^6$ interaction to study the properties of the structures. Furthermore, the $\phi^6$ theory allows for a connection with the models of high-energy physics, i. e., the ``bag models'' of Quarks within Hadrons \cite{Saxena}.

For the interaction (\ref{potential1}), the self-dual equations of topological structures are
\begin{align}\label{bps3}
    \frac{d\phi}{dx}=\pm\frac{1}{2}\chi^{n}\phi(1-\phi^2),
\end{align}
and 
\begin{align}\label{bps4}
    \frac{d\chi}{dx}=\pm\frac{1}{2}\chi(1-\chi^2).
\end{align}
Here, the parameter $n$ will be assumed to be an integer. In our model, the parameter $n$ is the degree of the generalization function. The modification of the parameter $n$ produces alterations of the structure in the topological sector of the field $\phi$. Let us mention that in Eq. (\ref{bps3}) it was considered that $f(\chi)=1/\chi^n$. This choice is one of the simplest forms of the $f(\chi)$ function that can induce contractions of the kink-like solutions of the $\phi$ field \cite{Bazeia1}.

For this particular model, the BPS energy density of the structure is
\begin{align}
    \mathcal{E}_{BPS}=\phi\phi'(1-\phi^2)+\chi\chi'(1-\chi^2),
\end{align}
i. e., 
\begin{align}\label{DE1}
    \mathcal{E}_{BPS}=\pm\frac{1}{2}[\chi^n\phi^2(1-\phi^2)^2+\chi^2(1-\chi^2)^2].
\end{align}

Let us now investigate the solutions of the BPS equations, i. e., Eqs. (\ref{bps3}) and (\ref{bps4}). For simplicity, in investigating the solutions of Eq. (\ref{bps4}), we have 
\begin{align}\label{soluçãochi}
    \int\,  \frac{d\chi}{\chi(1-\chi^2)}=\pm\frac{1}{2}\int\, dx.
\end{align}

\begin{figure}[ht!]
    \centering
    \includegraphics[height=7cm,width=7.3cm]{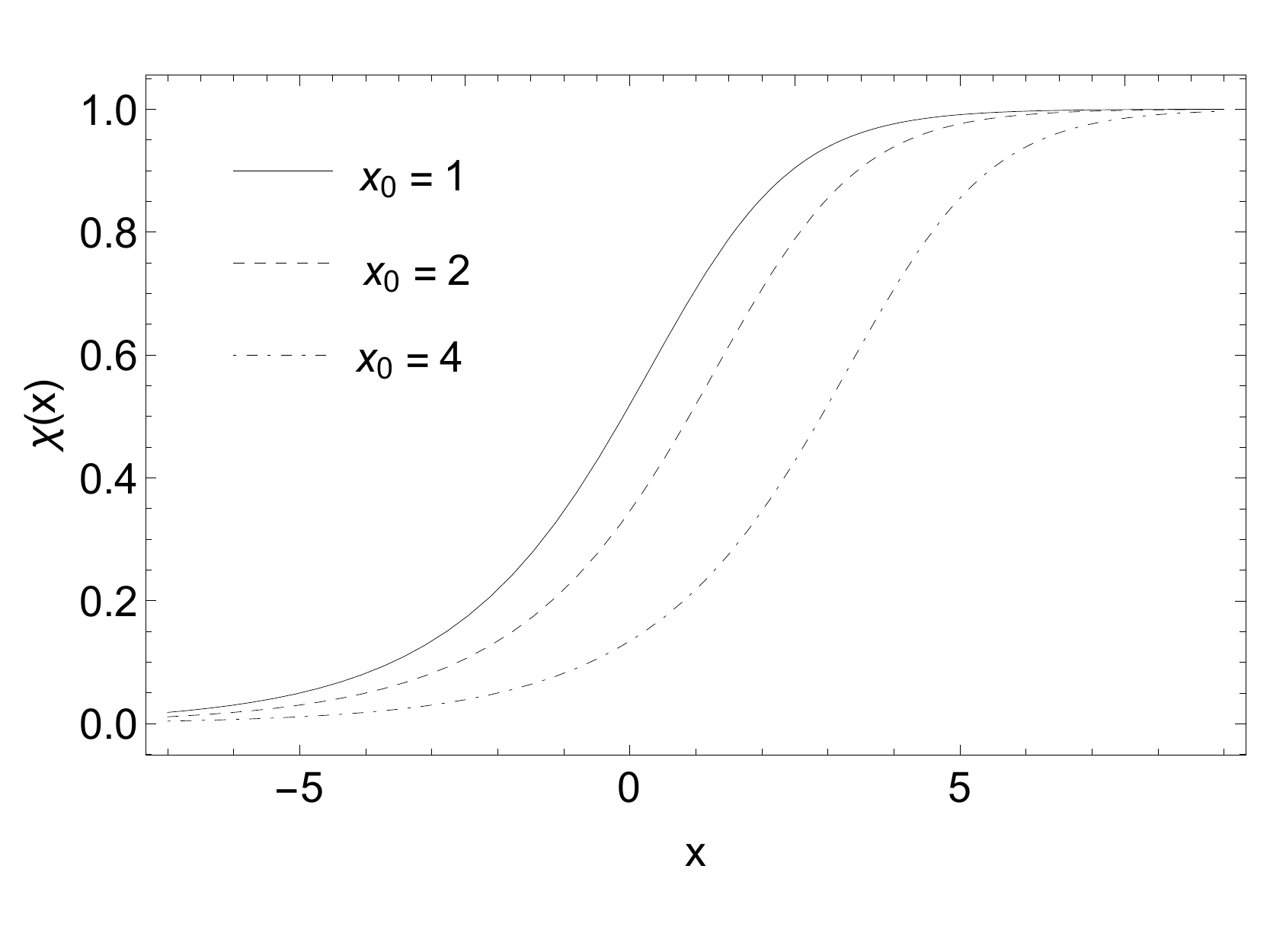}
    \includegraphics[height=7cm,width=7.3cm]{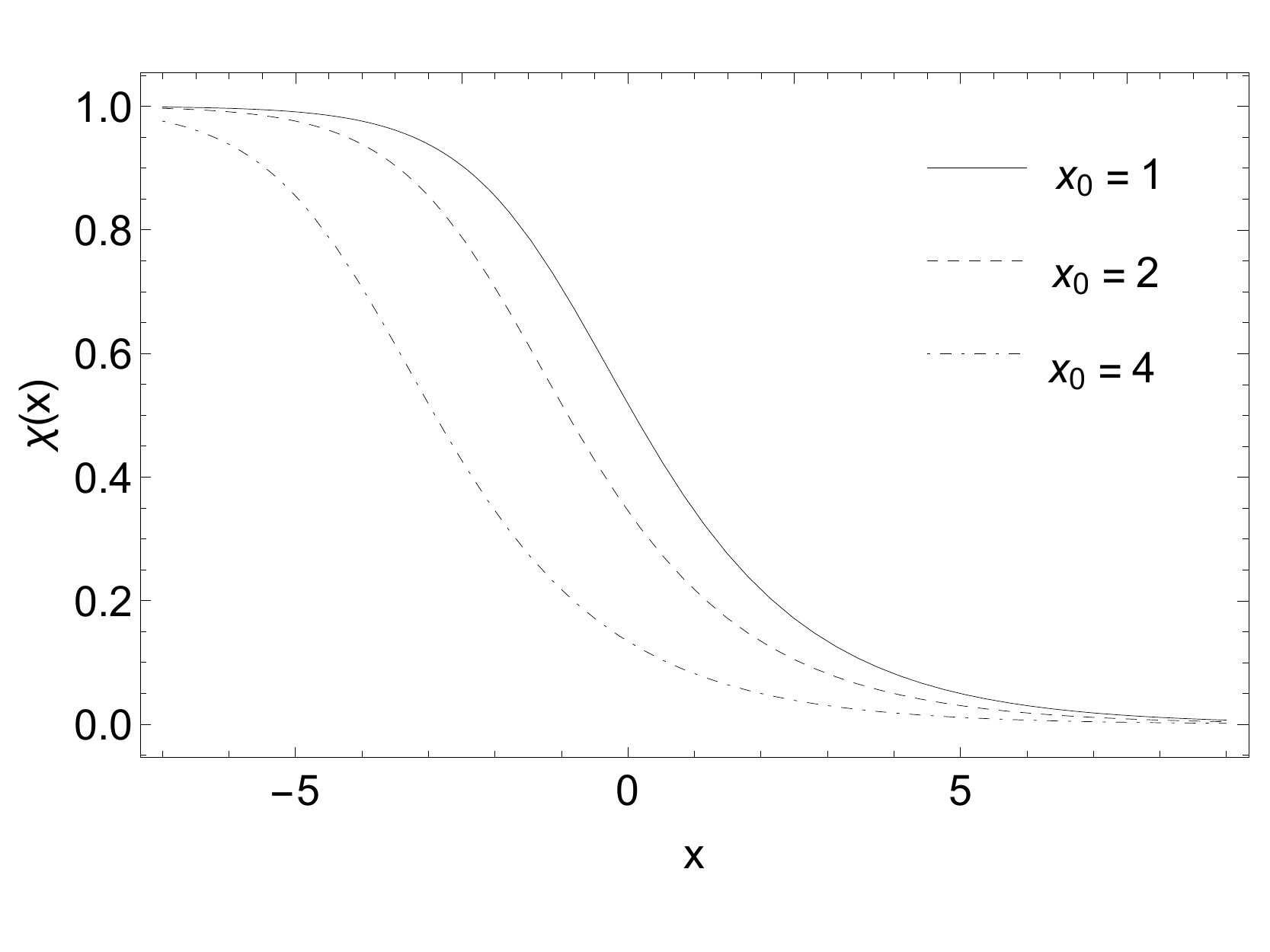}
    \vspace{-1cm}
    \begin{center}
        (a) \hspace{7cm} (b)
    \end{center}
    \vspace{-1cm}
    \caption{(a) Kink-like solutions of the $\chi(x)$ field when $x_0=1,$ $2$ and $4$. (b) Antikink-like solutions of the $\chi(x)$ field when $x_0=1,$ $2$ and $4$.}
    \label{fig1}
\end{figure}

The expression (\ref{soluçãochi}) leads us to kink and antikink solutions, namely,
\begin{align}\label{xsolution}
    \chi(x)=\frac{1}{\sqrt{1+\text{e}^{(x_0\mp x)}}}.
\end{align}
The negative sign of the expression above indicates kink-like solutions. On the other hand, the positive sign indicates the antikink-like solution. The value $x_0$ describes the starting position of the topological structure. In Fig. \ref{fig1}, the graphical behavior of the $\chi(x)$ field is displayed.

Using the solution (\ref{bps4}), one can find the analytical solution of the Eq. (\ref{bps3}). With the expression (\ref{xsolution}), we note that Eq. (\ref{bps3}) is rewritten as
\begin{align}\label{psolution}
    \frac{d\phi}{dx}=\pm\frac{1}{2[1+\text{e}^{(x_0\mp x)}]^{n/2}}\phi(1-\phi^2).
\end{align}

Rewriting the above expression, we arrive at
\begin{align}\label{psol}
    \int\, \frac{d\phi}{\phi(1-\phi^2)}=\pm\frac{1}{2}\int\, \frac{dx}{[1+\text{e}^{(x_0\mp x)}]^{n/2}}.
\end{align}

Let us remember that
\begin{align}\nonumber
    F(x,x_0)=&\frac{1}{2}\int\, \frac{dx}{[1+\text{e}^{(x_0\mp x)}]^{n/2}}\\ \nonumber
    =&\pm [1+\text{e}^{(\pm x_0-x)}]^{\mp n/2}[1+\text{e}^{(\mp x_0+x)}]^{\pm n/2}\,_{2}F_{1}\bigg(\frac{n}{2},\frac{n}{2},1+\frac{n}{2};\text{e}^{(-x_0\pm x)}\bigg).&
     \label{integrate}
\end{align}
Allow us to highlight that $_{2}F_{1}(a, b, c; y)$ is the well-known hypergeometric function. Thus, we define the hypergeometric function as
\begin{align}
    _2F_1(a,b,c;y)=\sum_{j=0}^{\infty}\frac{(a)_j(b)_j}{(c)_j}\frac{y^j}{j!}
\end{align}
where $a=\frac{n}{2}$, $b=\frac{n}{2}$, $c=\frac{n}{2}+1$, and $y=\text{e}^{(-x_0\pm x)}$.

\begin{figure}[ht!]
    \centering
    \includegraphics[height=7cm,width=7.3cm]{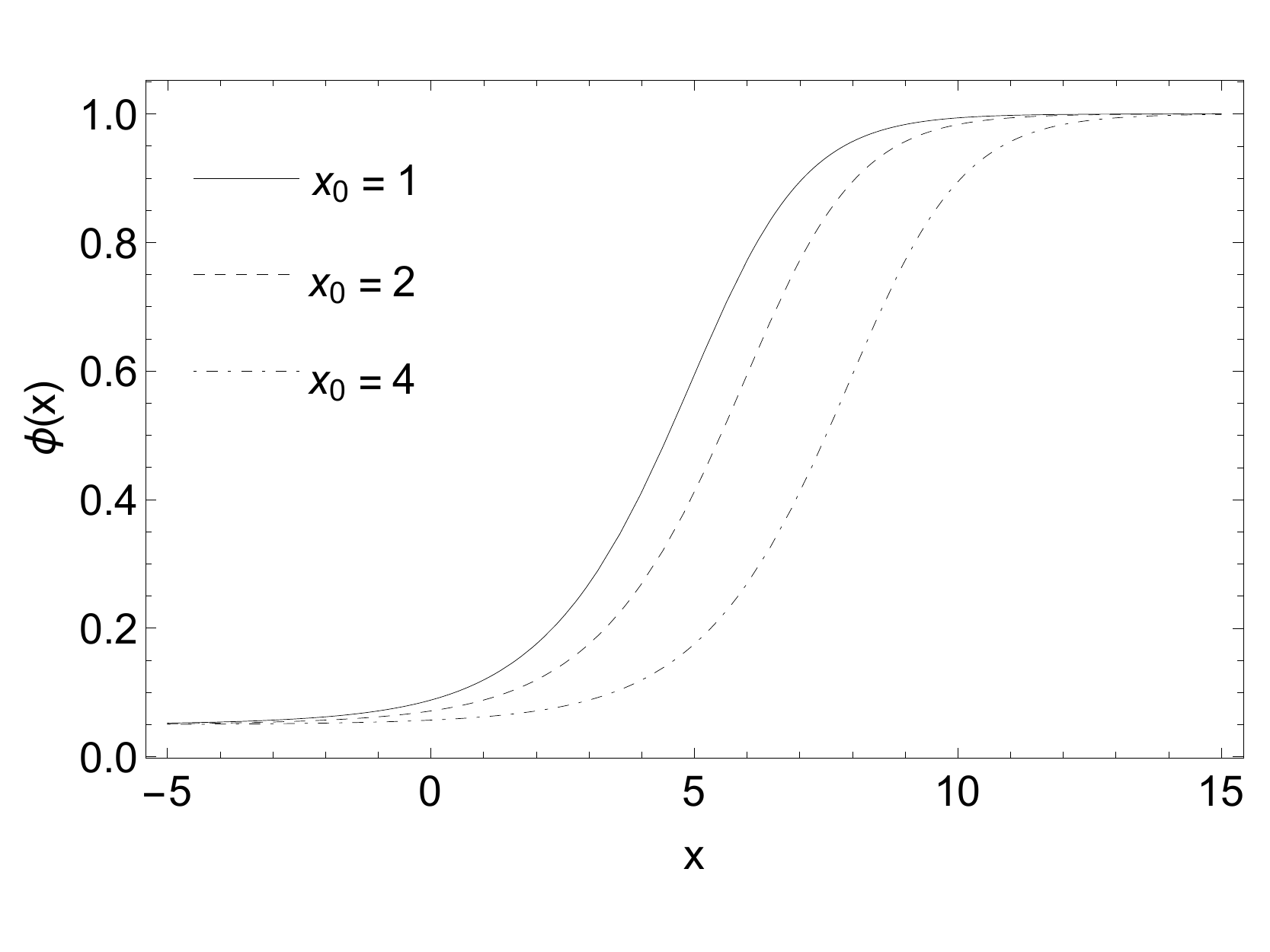}
    \includegraphics[height=7cm,width=7.3cm]{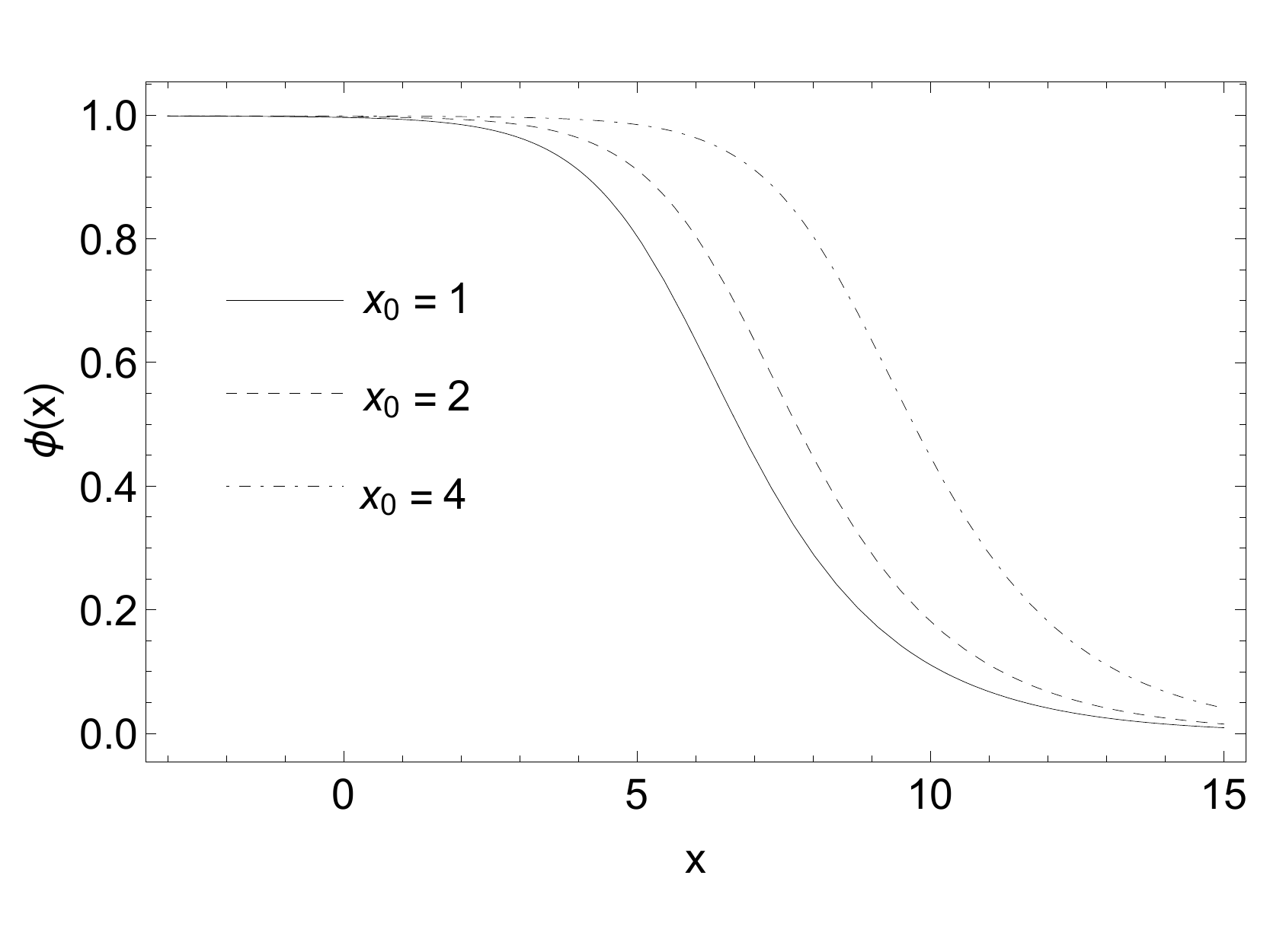}
    \vspace{-1cm}
    \begin{center}
        (a) \hspace{7cm} (b)
    \end{center}
    \vspace{-1cm}
    \caption{(a) Kink-like solutions of the $\phi(x)$ field when $c_0=6$, $n=1$,and $x_0=1,$ $2,$ and $4$. (b) Antikink-like solutions of the $\phi(x)$ field when $c_0=-6$, $n=1$, and $x_0=1,$ $2,$ and $4$.}
    \label{fig2}
\end{figure}

\begin{figure}[ht!]
    \centering
    \includegraphics[height=7cm,width=7.3cm]{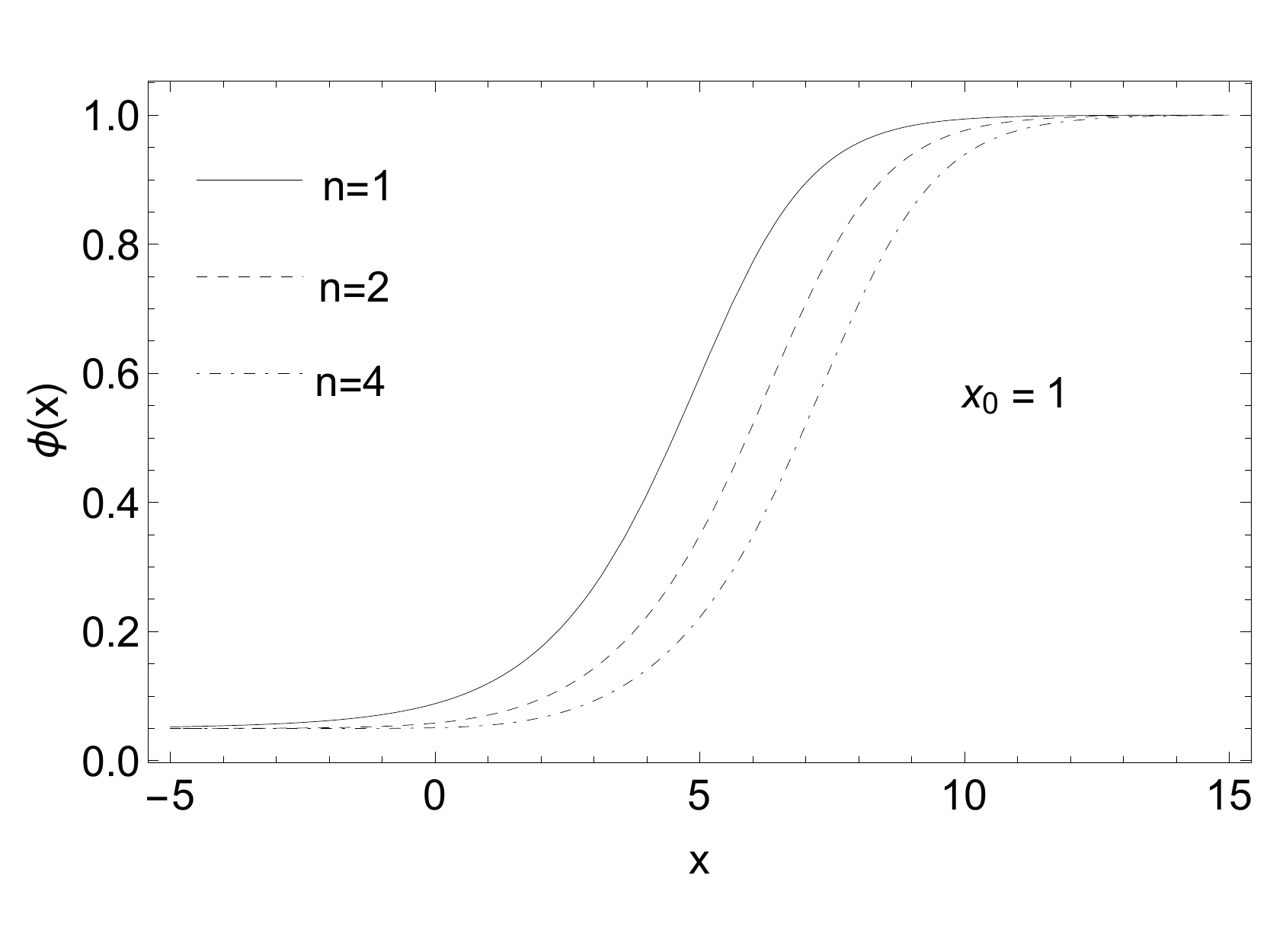}
    \includegraphics[height=7cm,width=7.3cm]{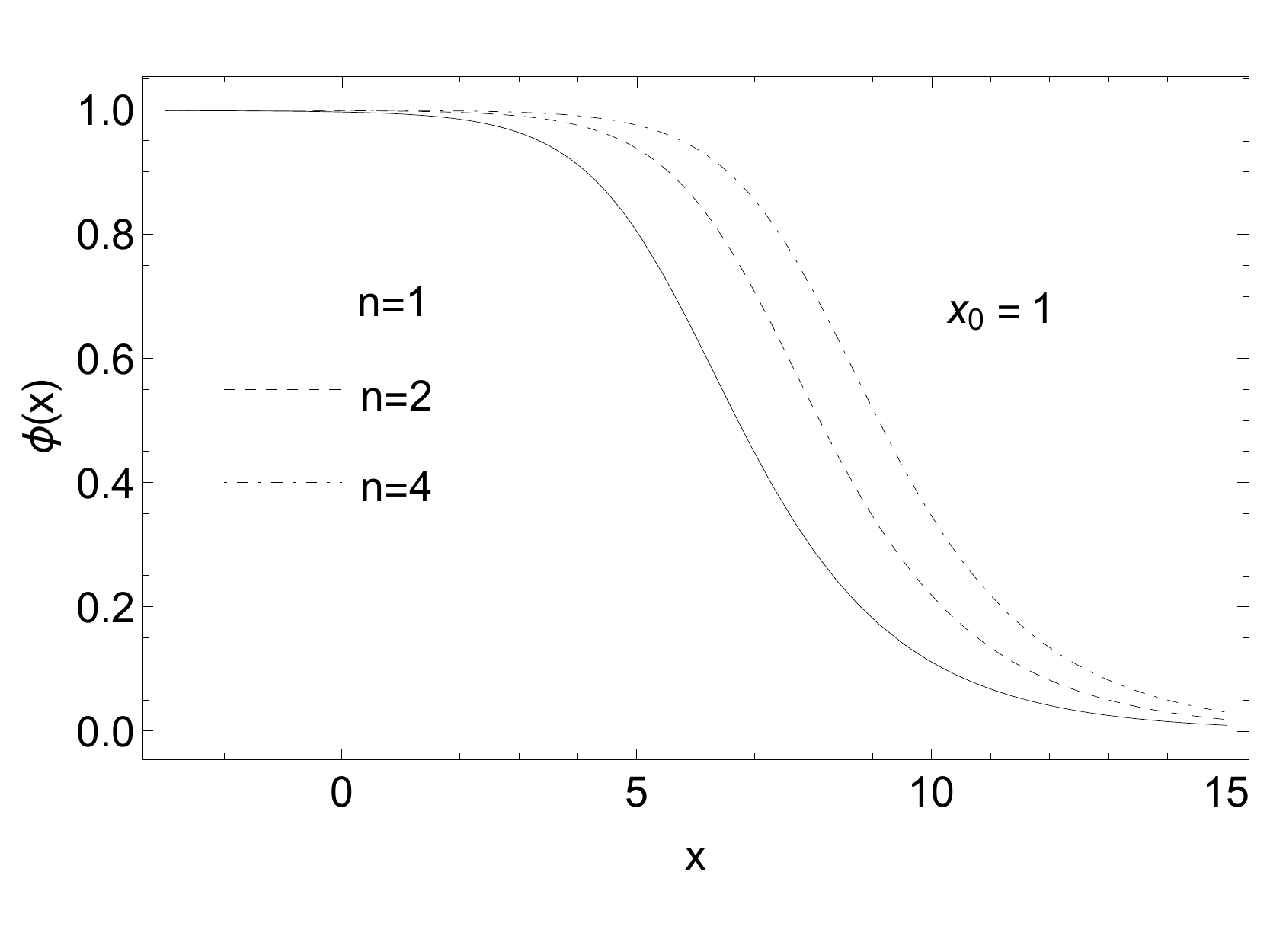}
    \vspace{-1cm}
    \begin{center}
        (a) \hspace{7cm} (b)
    \end{center}
    \vspace{-1cm}
    \caption{(a) Kink-like solutions of the field $\phi(x)$ when $x_0$ and $c_0$ are constant and $n$ varies. (b) Antikink-like solutions of the $\phi(x)$ field when $c_0$ and $x_0$ are constant and $n$ varies.}
    \label{fig3}
\end{figure}

Using (\ref{psol}) and (\ref{integrate}), solutions for the field $\phi$ are obtained. In this case, the solutions of the field $\phi$ are
\begin{align}\label{phisolution}
    \phi(x)=\frac{1}{\sqrt{1+\text{e}^{\mp 2F(x,x_0)\pm c_0}}}.
\end{align}
Here, $c_0$ is a constant of integration and will be responsible for the initial location of the structure in the topological sector of $\phi(x)$. The field configurations in the topological sector of $\phi$ are presented in Figs. \ref{fig2} and \ref{fig3}.

\begin{figure}[ht!]
    \centering
    \includegraphics[height=7cm,width=7.3cm]{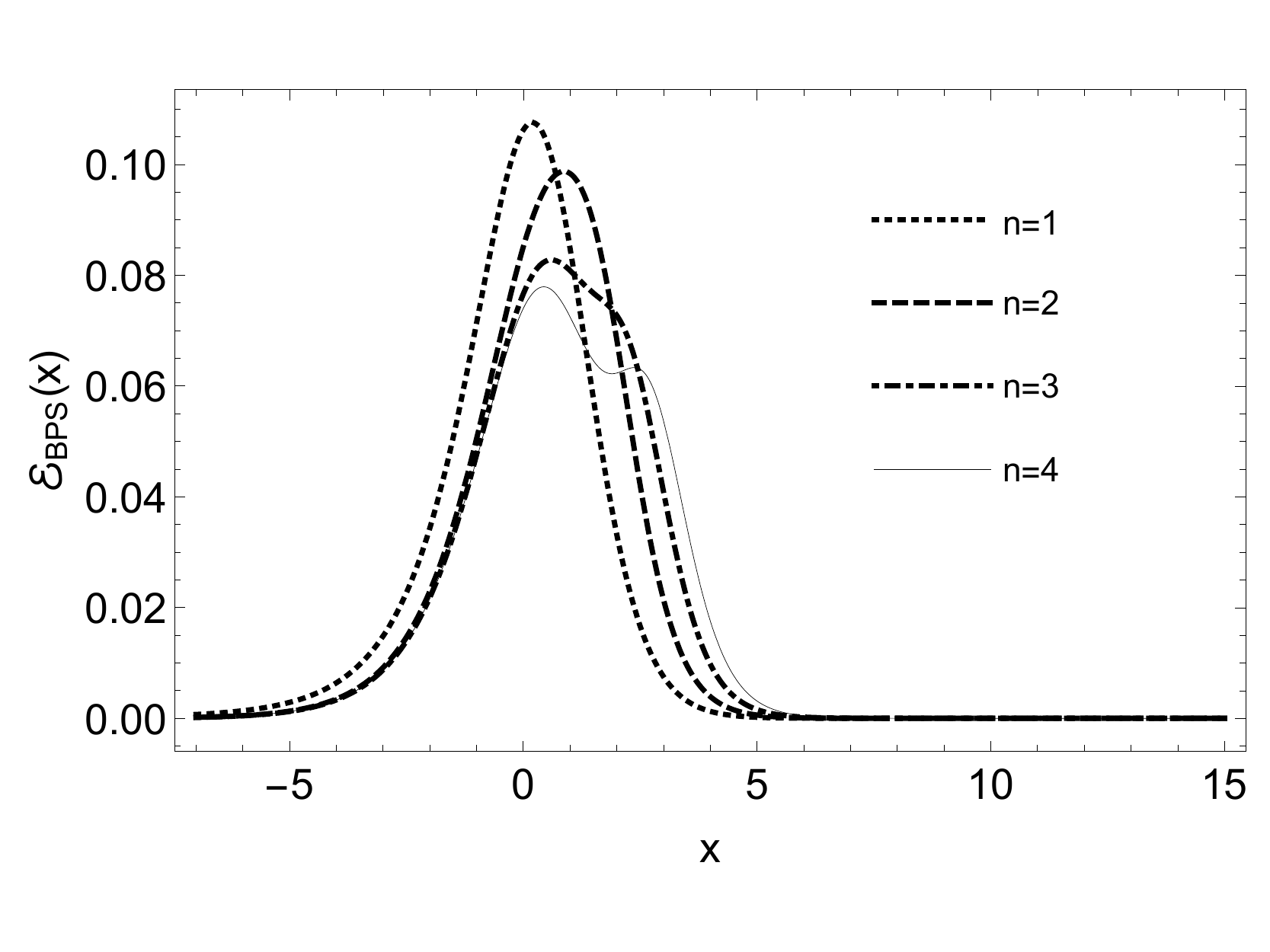}
    \vspace{-1cm}
    \caption{BPS energy density produced by fields $\chi$ and $\phi$.}
    \label{fig4}
\end{figure}

Substituting the Eqs. (\ref{xsolution}) and (\ref{phisolution}) in Eq. (\ref{DE1}), we obtain the BPS energy density. We display the BPS energy density profile of the model in Fig. \ref{fig4}. Note that when $n$ is small, the multi-field configurations of our theory begin to have an energy profile similar to true kinks, see Ref. \cite{Rajaraman}.

\section{Differential Configurational Entropy}

The results of the previous section suggest that the parameter n changes the topological sector structures $\phi$. This result allows the emergence of the following question: For which value of parameter n is the topological configurations found to be more stable? To answer this issue, we will use the concept of configurational entropy (or its variant, namely, DCE) to investigate the most likely field configuration.

Differential configurational entropy (DCE) is a variant of Configurational Entropy (CE). DCE plays a relevant role in measuring the informational complexity of a localized field configuration, see Refs. \cite{Gleiser,Wilami,MLA}. Furthermore, we can express the DCE as Fourier's transform of the energy density (BPS energy density).

To calculate the DCE is necessary to define it in terms of the probability density \cite{Gleiser2, Gleiser3, Gleiser4}. This probability density is
\begin{align}\label{densprob}
    f(\omega)=\frac{\vert \mathcal{F}(\omega)\vert^2}{\int_{-\infty}^{\infty}d\omega'\vert\mathcal{F}(\omega')\vert^2},
\end{align}
where
\begin{align}\label{fomega}
    \mathcal{F}(\omega)=-\frac{1}{\sqrt{2\pi}}\int_{-\infty}^{\infty}\mathcal{E}_{BPS}(x)\,\text{e}^{i\omega x}\,dx.
\end{align}

Thus, for a continuous and localized function, the DCE is given by
\begin{align}\label{dce}
    S_{C}[f]=-\int_{-\infty}^{\infty}f(\omega)\text{ln}[f(\omega)]\, d\omega.
\end{align}

Due to the profile of the fields $\phi(x)$ and $\chi(x)$, it is convenient to numerically analyze the DCE of the field configurations presented in Eqs. (\ref{xsolution}) and (\ref{phisolution}). To perform this analysis, we start by substituting the solutions (\ref{xsolution}) and (\ref{phisolution}) in the energy density expression, i. e., Eq. (\ref{DE1}). Substituting Eq. (\ref{DE1}) in (\ref{fomega}), you get $\mathcal{F(\omega)}$. Subsequently, applying the definition of probability density (\ref{densprob}) and DCE (\ref{dce}), respectively, we obtain the differential configurational entropy of the model previously discussed.

\begin{figure}[ht!]
    \centering
    \includegraphics[height=7cm,width=7.3cm]{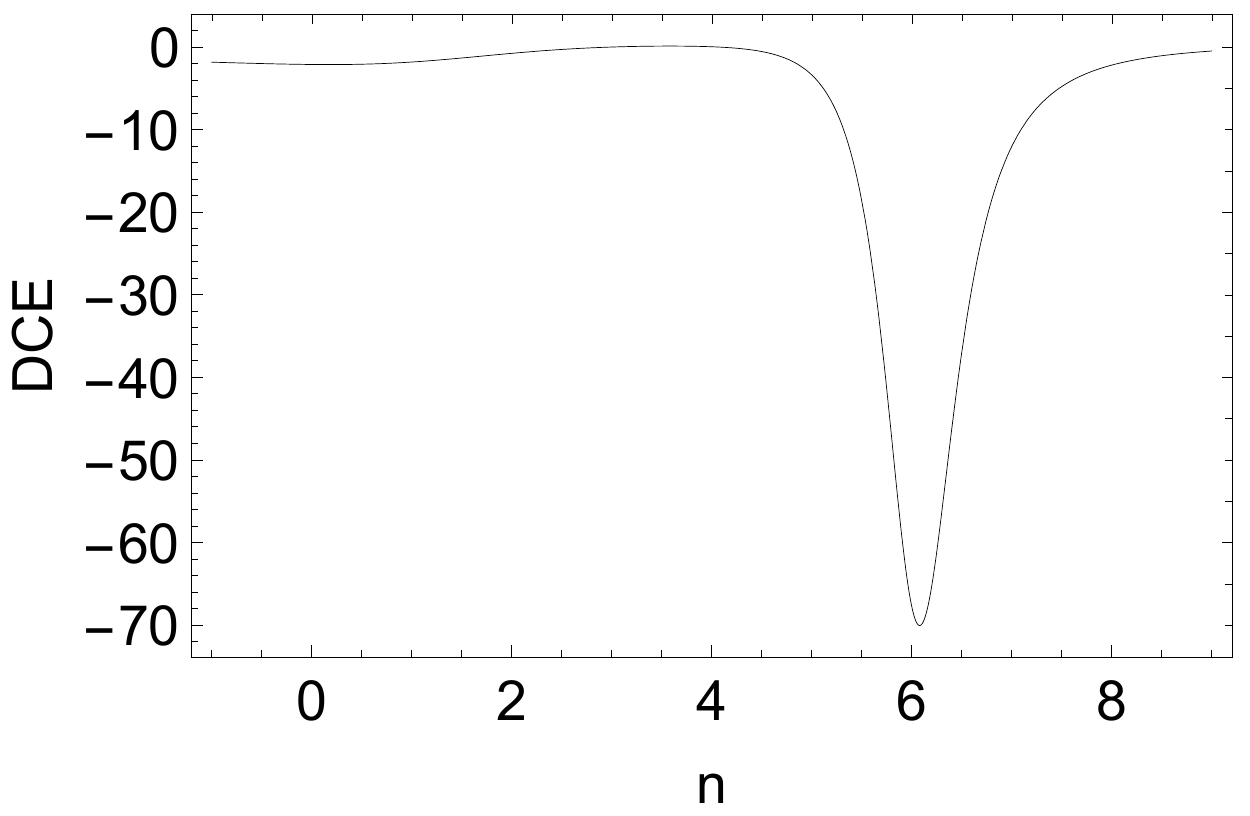}
    \vspace{-0.7cm}
    \caption{Differential configurational entropy in terms of the parameter $n$ when $c_0=x_0=1$.}
    \label{fig5}
\end{figure}

We display in Fig. \ref{fig5} the DCE result. By definition, the maximum or minimum point of the DCE indicates the most likely (or most stable) configurations, see Refs. \cite{LA2, Wilami, MLA}. Therefore, the structures most likely of our theory appear when the parameter $n\simeq 6$, as shown in Fig. \ref{fig5}. 

\begin{figure}[ht!]
    \centering
    \includegraphics[height=7cm,width=7.3cm]{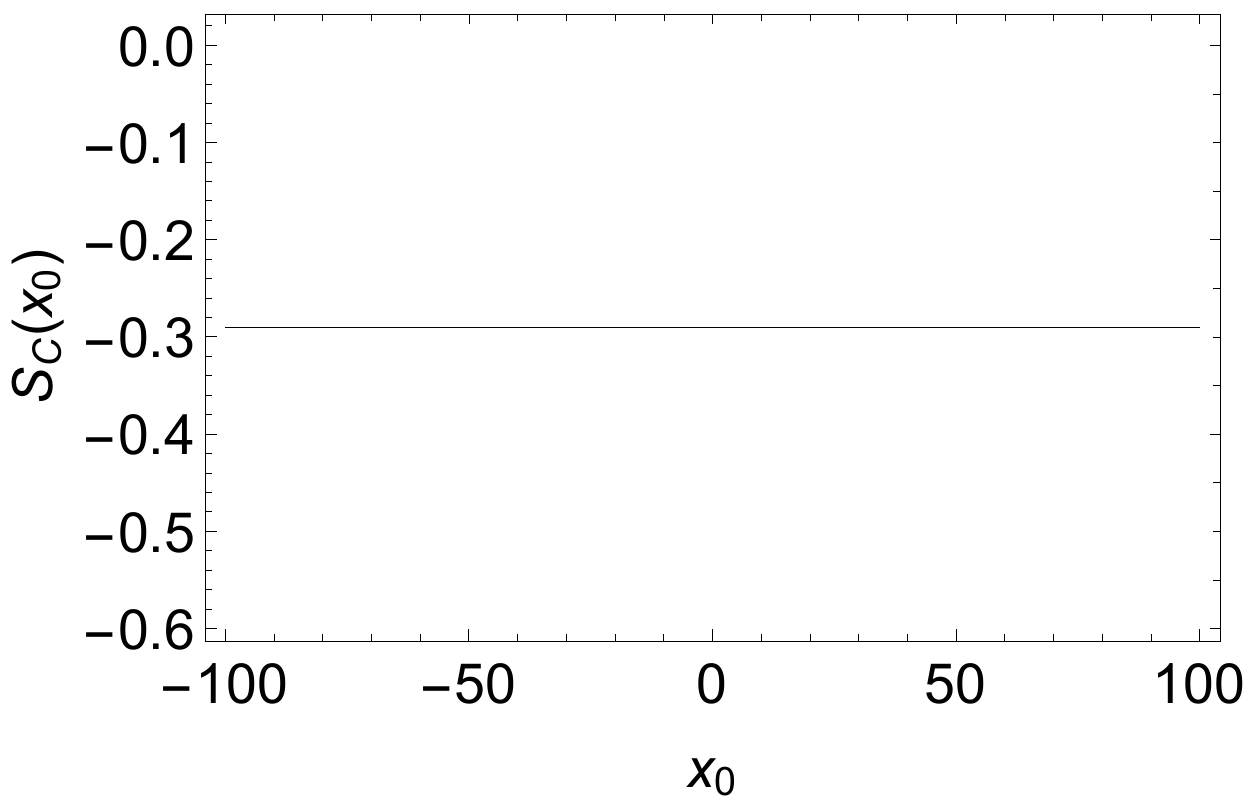}
    \includegraphics[height=7cm,width=7.3cm]{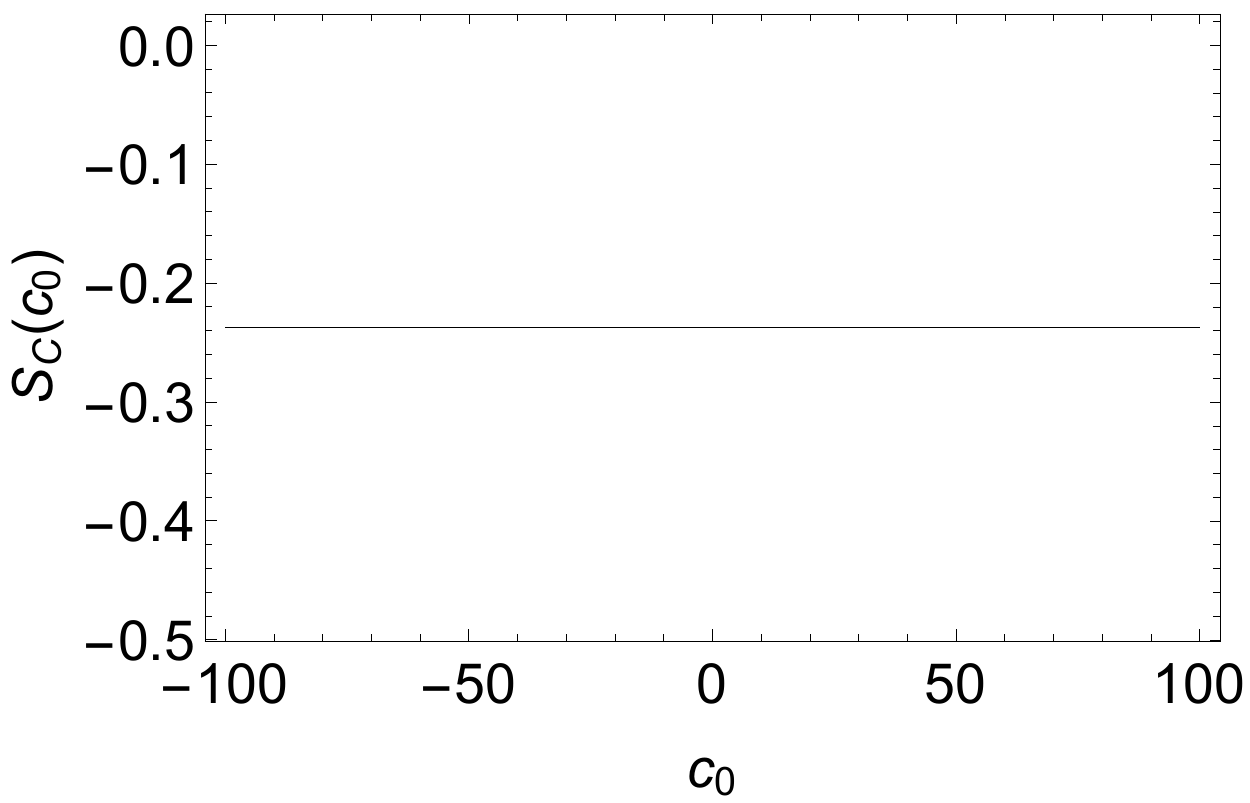}
    \vspace{-0.7cm}
    \begin{center}
    \, \, \,     (a) \hspace{7cm} (b)
    \end{center}
    \vspace{-1cm}
    \caption{Configurational differential entropy assuming n=1. (a) Varying the parameter $x_0$. (b) Varying the parameter $c_0$.}
    \label{fig66}
\end{figure}

Varying parameters $c_0$ and $x_0$ and keeping $n$ constant, i. e., $n=1$, one obtains the results of Fig \ref{fig66}. Fig. \ref{fig66} shows that DCE is constant when the parameters $c_0$ and $x_0$ vary. That occurs because these parameters describe the initial conditions of topological structures, and these configurations are equally likely. In other words, this is a consequence of having infinite initial configurations (i.e., regions where structures can be initially) for the system.

\section{Final remarks}

In this letter, we present a study of the field configurations of the $\phi^6$ theory with two scalar fields. Furthermore, to find the appropriate value for the parameter $n$, the concept of configurational entropy was used to search for the most stable and probable field configurations.

A peculiar result of our model is that it admits analytic solutions. Another result emerges when we look at the field setting. As a matter of fact, we show analytically that our theory has a contraction. We refer to contraction as the form that the fields only interpolate the vacuums  $\phi^{(0)}=0$ and $\phi^{(0)}=1$. Note the contrast to the usual $\lambda\phi^4$ theory, where we use the potential of type $V\propto (1-\phi^2)^2$ and the fields interpolate between the vacuums $\phi^{(0)}=\pm 1$.

\begin{figure}[ht!]
    \centering
    \includegraphics[height=7cm,width=7.3cm]{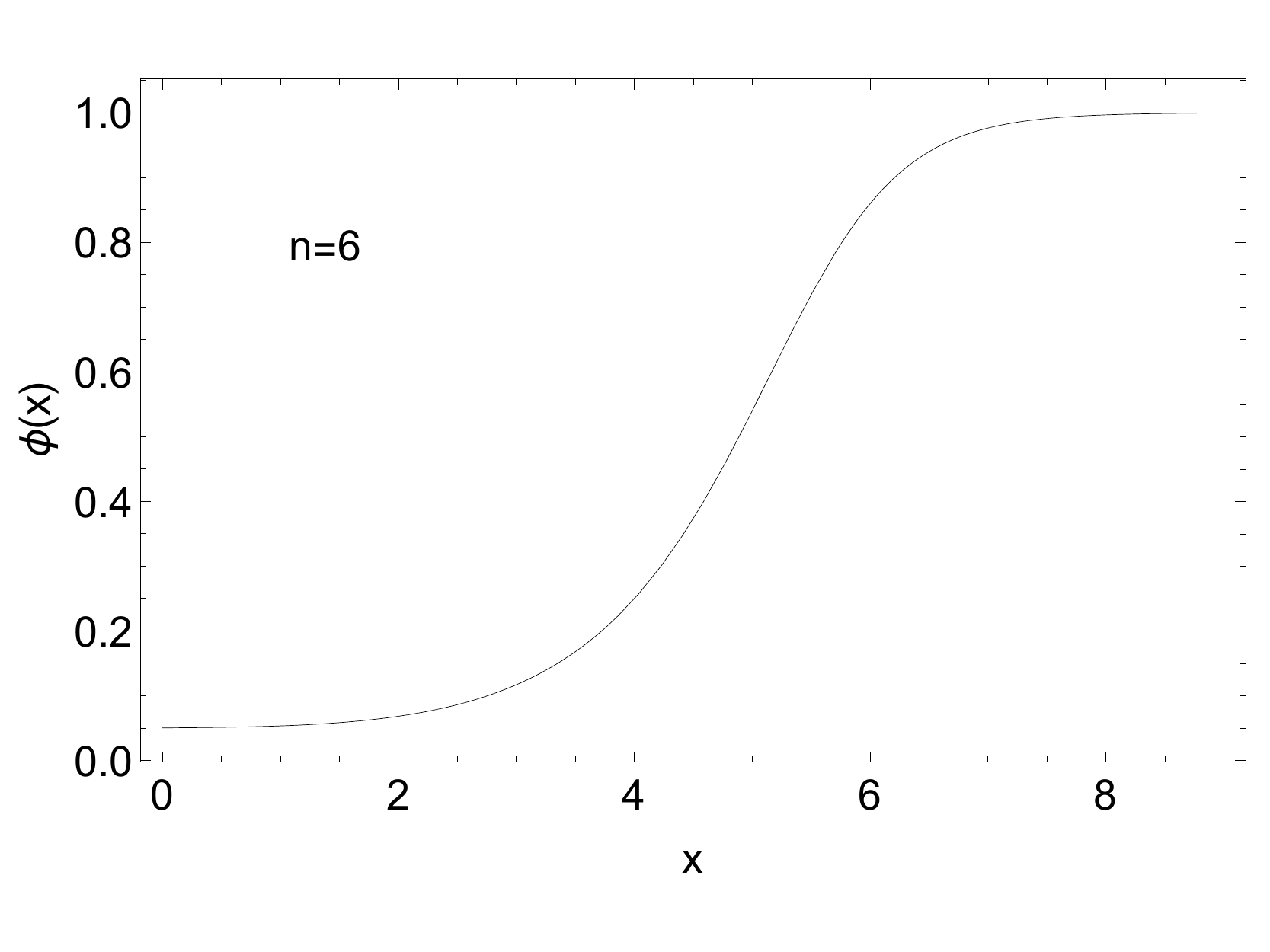}
    \includegraphics[height=7cm,width=7.3cm]{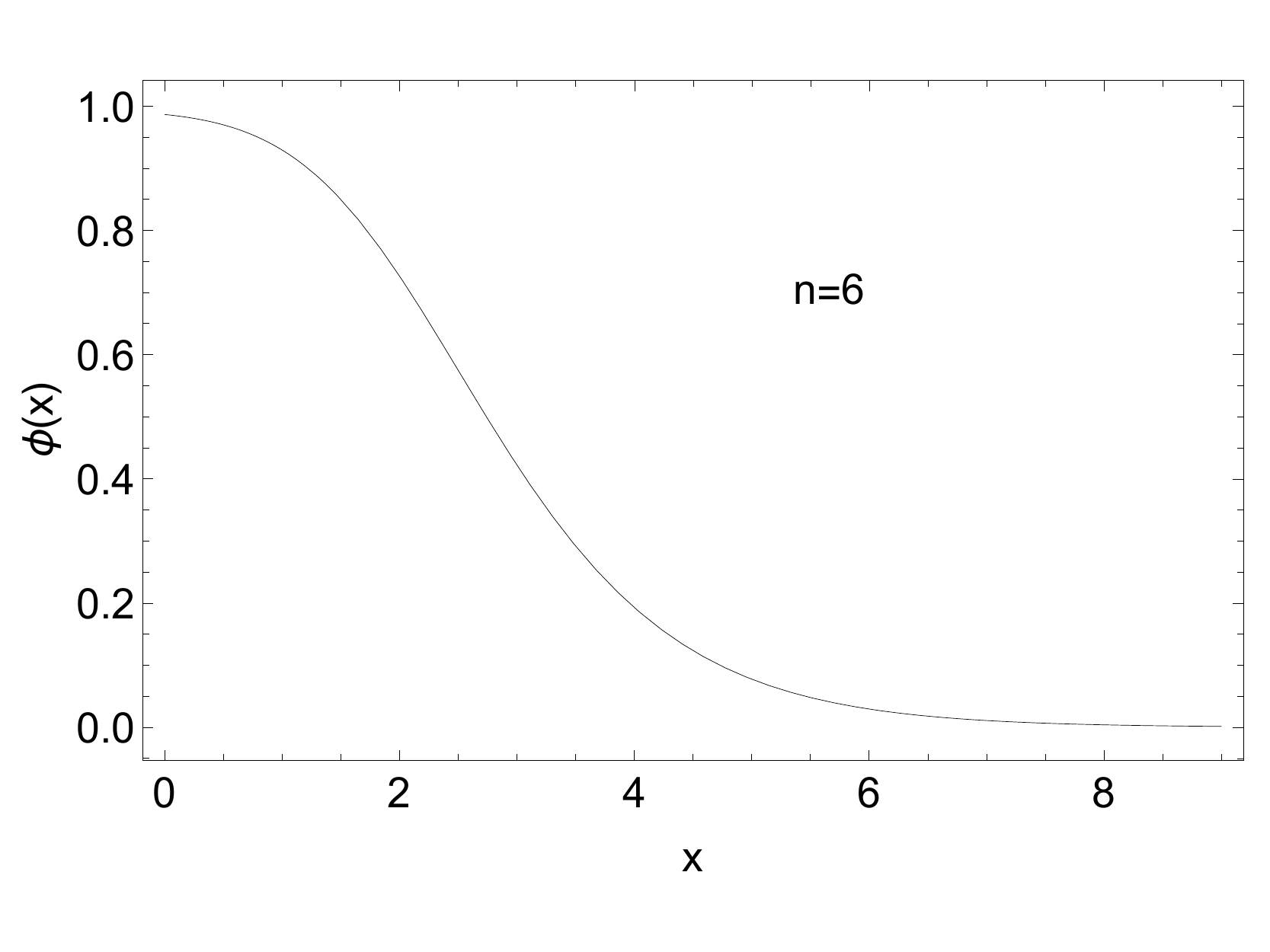}
    \vspace{-0.8cm}
    \caption{Most likely solutions based on DCE calculation.}
    \label{fig6}
\end{figure}

We found the appropriate value for the parameter n using the concepts derived from configurational entropy. Indeed, to reach our purpose, we use differential configurational entropy. The analysis of this quantity shows us numerically that the most likely, and therefore most stable, field configurations occur when $n=6$ approximately. Moreover, it is possible to see that for $n=6$, the field $\chi(x)$ will have the kink-like solutions shown in Fig. \ref{fig1}, while the field $\phi(x)$ have the kink-like configurations shown in Fig. \ref{fig6}.

Some future perspective of this work is the study of the scattering process in $\phi^6$ type theory with two or more coupled scalar fields. In addition, another perspective can be an investigation of the extension of this theory to models in which Lorentz symmetry breaking is considered. Not far away, a complementary result of this study can emerge by analyzing cases where a canonical scalar field and a noncanonical one are considered. The present work is the first step toward understanding these structures in a multi-field system. Therefore, we hope in the future to carry out studies in this directions to collaborate with the results presented here.

\section{Acknowledgments}
\hspace{0.5cm} The authors thank the Conselho Nacional de Desenvolvimento Cient\'{i}fico e Tecnol\'{o}gico (CNPq), grants n$\textsuperscript{\underline{\scriptsize o}}$ 309553/2021-0 (CASA) and the Coordena\c{c}\~{a}o de Aperfei\c{c}oamento de Pessoal de N\'{i}vel Superior (CAPES), grants n$\textsuperscript{\underline{\scriptsize o}}$ 88887.372425/2019-00 (FCEL), for financial support.

%%%%%%%%%%%%%%%%%%%%%%%%%%%%%%%%%%%%%%%%%%

\end{document}